\documentclass[preprint,onecolumn,superscriptaddress,longbibliography,amsmath,amssymb,aps]{revtex4-1}

\usepackage[pdftex]{graphicx}
\usepackage{amsmath}
\usepackage{bm}
\usepackage{xcolor}

\begin{document}

\title{Migration, trapping, and venting of gas in a soft granular material}

\author{Sungyon Lee}
\email{sungyon@umn.edu}
\affiliation{Department of Mechanical Engineering, University of Minnesota, Minneapolis, MN 55455, USA}
\author{Jeremy Lee}
\affiliation{Department of Engineering Science, University of Oxford, Parks Road, Oxford OX1 3PJ, UK}
\author{Robin Le Mestre}
\affiliation{D{\'e}partement g{\'e}nice m{\'e}canique, ENS Cachan, 94235 Cachan cedex, France}
\author{Feng Xu}
\affiliation{Department of Mechanical Engineering, University of Minnesota, Minneapolis, MN 55455, USA}
\author{Christopher W. MacMinn}
\email{christopher.macminn@eng.ox.ac.uk}
\affiliation{Department of Engineering Science, University of Oxford, Parks Road, Oxford OX1 3PJ, UK}

\date{\today}

\begin{abstract}
Gas migration through a soft granular material involves a strong coupling between the motion of the gas and the deformation of the material. This process is relevant to a variety of natural phenomena, such as gas venting from sediments and gas exsolution from magma. Here, we study this process experimentally by injecting air into a quasi-2D packing of soft particles and measuring the morphology of the air as it invades and then rises due to buoyancy. We systematically increase the confining pre-stress in the packing by compressing it with a fluid-permeable piston, leading to a gradual transition in migration regime from fluidization to pathway opening to pore invasion. We find that mixed migration regimes emerge at intermediate confinement due to the spontaneous formation of a compaction layer at the top of the flow cell. By connecting these migration mechanisms with macroscopic invasion, trapping, and venting, we show that mixed regimes enable a sharp increase in the average amount of gas trapped within the packing, as well as much larger venting events. Our results suggest that the relationship between invasion, trapping, and venting could be controlled by modulating the confining stress.
\end{abstract}

\maketitle

\section{Introduction}

The buoyancy-driven migration of gas through a liquid-saturated granular material is central to a variety of geophysical processes, from the generation and venting of biogenic gases from sediments~\cite{boudreau2005, jain2009, scandella2011} to the exsolution and venting of volatile gases from magma~\cite{suckale2010, oppenheimer2015, parmigiani2016}. The former process is important for aquatic ecosystems and the latter contributes to the strength of volcanic eruptions; both play an important role in the global carbon cycle~\cite{skarke2014, parmigiani2016}.

The grain-scale mechanics of gas migration and venting are relatively well understood. Essentially, the gas migrates in one of two ways: (1) By invading the pore space between the solid grains (displacing the liquid), or (2) by propagating macroscopic pathways or cavities through the packing (displacing the liquid-grain mixture). From these two basic migration mechanisms, a wide variety of macroscopic migration patterns can emerge, depending on the relative strengths of the forces driving migration (injection, buoyancy), the forces resisting grain motion (friction, cohesion, elasticity, confining stress), and the force resisting the entry of gas into the pore space (capillarity).

Previous work has focused almost exclusively on pattern formation during the injection of gas into quasi-2D packings of dense, rigid, non-cohesive particles (glass beads or sand). For example, gas invasion in horizontal systems has been studied extensively~\cite{chevalier2007, chevalier2009, sandnes2007, sandnes2011, holtzman2012, eriksen2015}. In these systems, gas invasion is driven by injection and particle motion is resisted by friction. These systems are ideal for (and also limited to) studying pattern formation during initial gas invasion, because the particle motion is irreversible and the invasion pattern is fixed once it is formed. Initial invasion and steady injection have also been studied extensively in the context of vertical systems, where gas invasion is driven by both injection and buoyancy and particle motion is resisted by both friction and confining stress~\cite{varas2009, varas2011a, varas2011b, varas2013, varas2015, dalbe2018, barth2019, sun-jgr-2019}. These systems allow for the possibility of studying continuous gas injection because the associated migration patterns are transient, with the confining stress forcing cavities and pathways to close and disconnect after gas venting.

While fascinating, migration patterns are typically an intermediate step between grain-scale mechanics and the macro-scale dynamics of gas trapping and venting. The latter topic has received less attention, and many fundamental questions remain about the characteristic size and frequency of venting events, the amount of gas trapped within the medium, and the relationship between the injection rate and the venting rate.

Here, we connect migration mechanisms and patterns with the macro-scale dynamics of gas trapping during the injection and buoyancy-driven migration of gas through a monolayer of hydrogel particles. Relative to glass beads or sand, these particles are soft, slippery, and nearly neutrally buoyant. These differences have three important implications for the mechanics of the packing:
\begin{itemize}
    \item \textbf{Elasticity:} Packings of stiff particles deform in a way that is macroscopically irreversible because the particles store a negligible amount of elastic energy, so macroscopic cavities and pathways will remain open once formed. Our packings are macroscopically elastic due to the elasticity of the particles themselves, such that cavities and pathways tend to close~\cite{macminn2015}. Muddy sediments comprise water, mineral particles, organic matter, and numerous small, trapped gas bubbles; the latter two constituents lead to macroscopic elasticity despite the stiffness of the mineral particles~\cite{boudreau2005}.
    \item \textbf{Friction:} The sliding and bridging of frictional particles against confining walls strongly localizes deformations, preventing forces and displacements from propagating more than a few gap-thicknesses in any direction. These wall effects are particularly prominent in quasi-2D systems; for example, the mechanics of so-called frictional-fluid systems are controlled almost entirely by particle-wall friction~\cite{sandnes2011}. The low sliding friction of our hydrogel particles enables long-range mechanical interactions, allowing forces and displacements to span the full extent of the flow cell~\cite{macminn2015}. However, the lack of friction or cohesion between the particles allows them to rearrange more readily than might be expected in a natural system.
    \item \textbf{Gravity:} In the presence of gravity, the effective stress within a saturated packing of negatively buoyant grains will increase with depth. For a packing of glass beads or sand in water, this implies a variation in effective stress of nearly $\sim$$1\,\mathrm{kPa}$ over a distance of $100\,\mathrm{mm}$; in our packings, this variation is $\sim$$3\,\mathrm{Pa}$ over the same vertical distance.

For frictional particles in a confined system, the vertical stress profile is further complicated by the Janssen effect, where the vertical walls support some fraction of the particle weight~\cite{sun-jgr-2019}. In our system, we expect the lack of sliding friction and near-neutral buoyancy to provide a nearly uniform stress profile. This profile is inappropriate for sediment, where confining stress is indeed expected to increase with depth, but also much simpler than that of dense, frictional grains. Note that negative particle buoyancy also provides a mechanism for reversible energy storage, and therefore also a restoring force that would tend to close cavities and pathways; this gravitational restoring force is negligible in our system relative to the elastic restoring force.
\end{itemize}

The softness and slipperiness of our particles also enable working with a quasi-2D monolayer, which is difficult to achieve with glass beads or sand. This monolayer allows for clear visualisation and quantitative analysis of the gas distribution, including both gas within the pore space and gas in macroscopic cavities and pathways. By varying the initial confining stress, we explore the full range of possible migration regimes in a single experimental system while simultaneously visualizing both gas distribution and grain motion. We focus in particular on the transition between migration regimes, showing that different mechanisms can co-exist and interact within the flow cell in a nontrivial way, leading to strongly non-monotonic variation in gas trapping and venting with confinement.

\section{Experimental system}\label{sec:setup}

\begin{figure}[t]
    \centering
    \includegraphics[width=17.2cm]{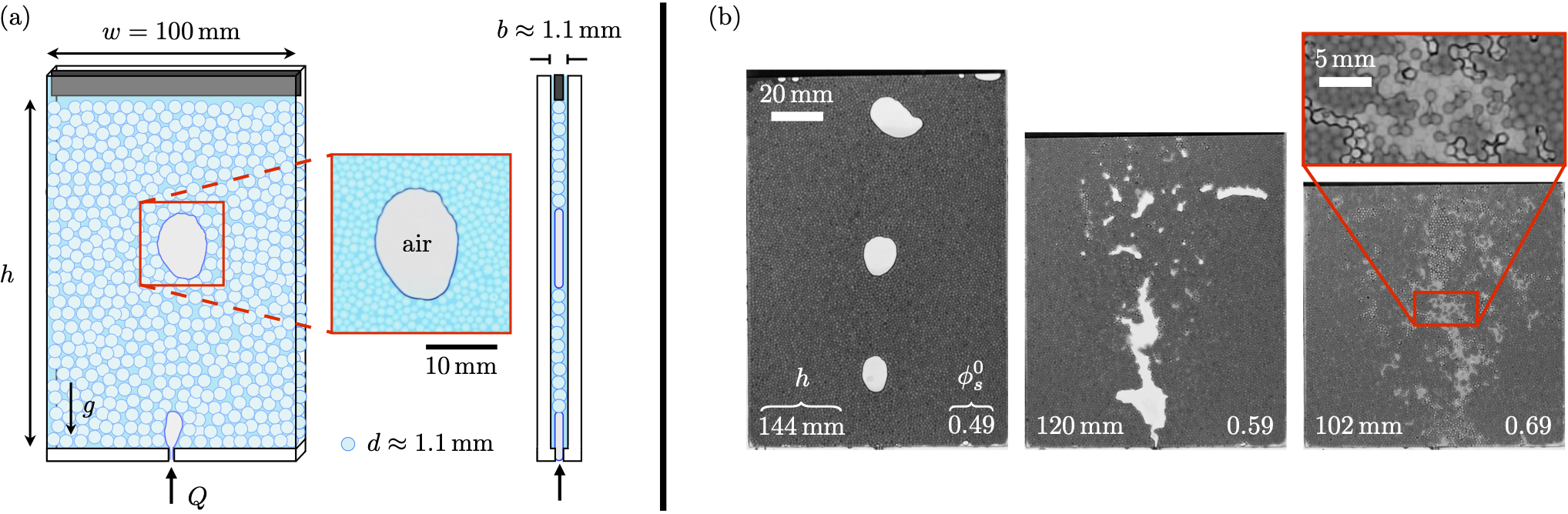}
    \caption{(a) We inject air into the bottom of a vertical flow cell filled with a liquid-saturated monolayer of soft particles. The flow area is formed by two glass plates clamped to a rigid spacer, with adjustable height $h$ set by a fluid-permeable piston at the top. We control the initial solid fraction $\phi_s^0$ and therefore also the initial confining effective stress $|\sigma_0^\prime|$ by changing $h$. After fixing $\phi_s^0$, we inject air with a syringe pump and then record the dynamics of air invasion, migration, trapping, and venting, as well as the deformation of the packing, at high resolution using a digital camera. (b) Snapshots at three representative values of $\phi_s^0$ show that under weak confinement (left), discrete gas bubbles rise by fluidizing the packing; under moderate confinement (middle), gas rises by opening transient, slender, fracture-like pathways; and under strong confinement (right), gas rises by invading the pore space between the grains and the walls. \label{fig:experiments} }
\end{figure}

Our experimental system consists of a vertical flow cell filled with $N\approx{}11,000$ spherical beads of polyacrylamide hydrogel (JRM Chemical; Fig.~\ref{fig:experiments}a). The packing is initially fully saturated with a mixture of water and glycerol (67\% glycerol by mass), and each bead is a cross-linked network of polymer chains that is saturated with the same fluid. The volume fraction of polymer within each bead is less than 1\%, so the beads are soft and elastic (Young modulus of tens of $\mathrm{kPa}$), nearly neutrally buoyant in the liquid~(denser by a few $\mathrm{kg}\,\mathrm{m}^{-3}$), and strongly wetted by the liquid relative to air. The beads are also extremely slippery~(friction coefficient of $\sim$$10^{-2}$)~\cite{mukhopadhyay-pre-2011, panaitescu-pre-2017, jewel2018}. The beads have mean diameter $d\approx1.1\,\mathrm{mm}$ with about 10\% polydispersity. The flow area has fixed gap thickness $b=1.26\,\mathrm{mm}$, width $w=100\,\mathrm{mm}$, and variable height $h$. Choosing the gap thickness to match the bead diameter produces a quasi-2D monolayer of particles, enabling individual particle tracking~\cite{macminn2015}.

In the series of experiments reported here, we explore the impact of confinement on gas migration, trapping, and venting. To do so, we vary the initial compressive confining effective stress $\sigma^\prime_0<0$ while keeping all other parameters fixed. We vary $\sigma^\prime_0$ in our system by compressing the packing to different fixed heights $h$ with a fluid-permeable piston, which imposes an initial solid fraction $\phi_s^0\approx{}\pi{}d^3N/(6wbh)$. Note that this value is a volumetric solid fraction for a planar packing of spheres confined between two plates, not an areal solid fraction for a packing of circles. For a frictionless granular packing, we expect that $|\sigma^\prime_0|=0$ for $\phi_s^0<\phi_s^\star$, where $\phi_s^\star$ is the jamming threshold, and that $|\sigma^\prime_0|$ is a monotonically increasing function of $\phi_s^0$ for $\phi_s^0>\phi_s^\star$. For simplicity, we take $\phi_s^\star$ to be the random loose packing fraction that results from allowing the particles to settle slowly under gravity, $\phi_s^\star\approx{0.51}$ (the particles are dense enough to settle and large enough to be non-Brownian, but the resulting ``lithostatic'' effective stress within the packing is only a few $\mathrm{Pa}$). Note that, during each experiment, the value of solid fraction $\phi_s$ (and therefore also of $\sigma^\prime(\phi_s)$) will vary locally in space and time due to the migrating air.

Prior to each experiment, we remove the piston entirely and then inject air into the unconfined packing at a high rate in order to vigorously rearrange the particles and also to release trapped air from prior experiments. Once all of the air has escaped, we allow the particles to settle and then replace the piston. To perform an experiment, we fix $h$ to a value between $99$ and $153\,\mathrm{mm}$, and inject air into the bottom of the flow cell at a constant nominal rate $Q=3\,\mathrm{mL}/\mathrm{min}$ with a syringe pump (New~Era~NE-4000) for 10 minutes. The air is compressible, so the actual injection rate varies in time. We measure the injection pressure $p_g^\mathrm{inj}$ throughout the experiment with a pressure transducer (Honeywell~40PC001B1A). We would expect a constant injection pressure to lead to qualitatively similar results. We image the cell at moderate frame rate (30\,fps) and high spatial resolution ($\sim$$8\,\mathrm{px}\,\mathrm{mm}^{-1}$) with a digital camera, allowing for quantitative measurement and analysis of migration, trapping, and deformation.

\section{Results}

\subsection{Air migration regimes}\label{ss:regimes}

Varying $\phi_s^0$ from 0.46 to 0.71 spans the full expected range of migration regimes, from fluidization under weak confinement to pathway opening (``fracturing'') under moderate confinement to pore invasion under strong confinement (Fig.~\ref{fig:experiments}b). Pore invasion is the entry of the air into the pore space between the grains by displacing the liquid. In contrast, both pathway opening and fluidization involve the creation and propagation of macroscopic cavities in the packing by displacing both liquid and grains. In pathway opening, these cavities are slender and fracture-like. In fluidization, these cavities are round and bubble-like.

\begin{figure*}
    \centering
    \includegraphics[width=17.2cm]{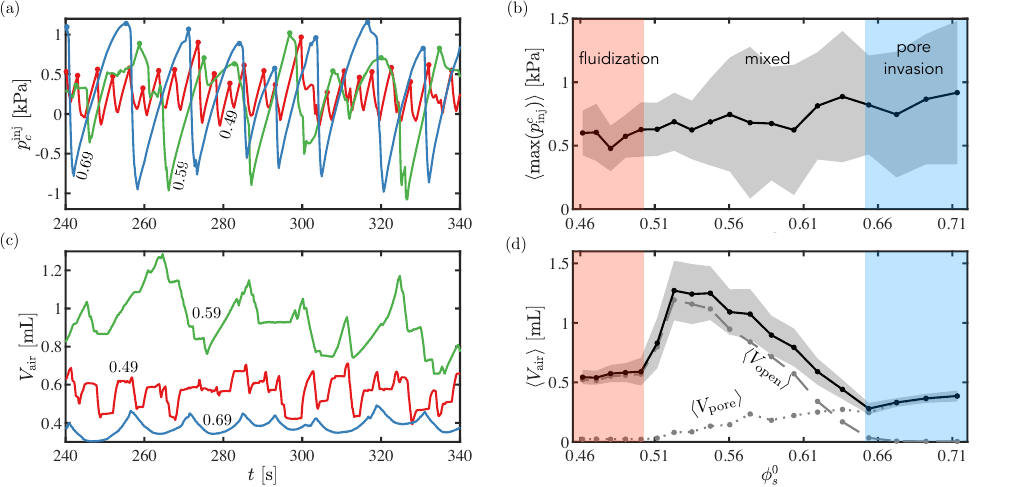}
    \caption{(a)~We show the injection capillary pressure $p_c^\mathrm{inj}$ as a function of time $t$ for $\phi_s^0=0.49$ (red), $0.59$ (green), and $0.69$ (blue) (dots indicate local peaks). (b)~The average peak value $\langle \mathrm{max}(p_c^\mathrm{inj})\rangle$ is noisy, but increases relatively steadily and monotonically with $\phi_s^0$. (c)~The total volume of air in the cell $V_\mathrm{air}$ is also oscillatory in all cases, but the mean value and the amplitude and period of the oscillations are distinctly larger in the intermediate mixed regimes than in either of the two end-member regimes. (d)~The time-averaged total air volume $\langle{}V_\mathrm{air}\rangle$ (solid line) exhibits a strongly non-monotonic dependence on $\phi_s^0$, increasing sharply at $\phi_s^0=0.52$ and then falling gradually. $\langle V_\mathrm{air}\rangle$ can be divided into two contributions: The amount of air trapped inside the pore space $\langle V_\mathrm{pore}\rangle$ (dotted line) and the amount of air in open bubbles and pathways $\langle V_\mathrm{open}\rangle$ (dashed line), where $\langle V_\mathrm{air}\rangle=\langle V_\mathrm{pore}\rangle+\langle V_\mathrm{open}\rangle$. In (b) and (d), the black dots and line show the mean and the gray band has a width of two standard deviations. For $\phi_s^0\geq0.59$, each result is the average of at least two experiments. \label{fig:trapping} }
\end{figure*}

To rationalize these observations, we consider the injection capillary pressure $p_c^\mathrm{inj}$, which is the difference between the injection pressure (in the gas) and the hydrostatic pressure at the bottom of the flow cell (in the liquid), $p_c^\mathrm{inj}=p_g^\mathrm{inj}-\rho_wgh_l$. The height $h_l$ of the liquid varies in time because gas invasion displaces liquid from the flow region, but these variations are only a few percent of $h_l$ and it is always the case that $h_l>h$. For gas to enter the flow cell, $p_c^\mathrm{inj}$ must exceed the smallest of three threshold pressures: (1)~the capillary pressure $p_c^y$ needed to inflate a cavity by yielding the packing, (2)~the capillary pressure $p_c^\mathrm{frac}$ needed to open a pathway in the packing, and (3)~the capillary pressure $p_c^e$ needed to enter the pore space (the capillary entry pressure). The capillary pressure must also exceed the entry pressure of the injection port itself (0.1--0.2~kPa), but this threshold is lower than the other three except at vanishing confinement (cf. Fig.~\ref{fig:trapping}a,b).

The relative sizes of these three threshold pressures depend on $|\sigma^\prime_0|$. We estimate the yield pressure as $p_c^y\sim\sigma_y+|\sigma^\prime_0|$, where $\sigma_y$ is the yield stress of the packing. A saturated packing of hydrogel particles is a non-cohesive granular material for which we expect $\sigma_y\sim\alpha|\sigma^\prime_0|$ with $0<\alpha\lesssim{}1$, so that $p_c^y\sim(1+\alpha)|\sigma^\prime_0|$~\citep[\textit{e.g.},][]{jewel2018}. We estimate the fracture pressure as $p_c^\mathrm{frac}\sim{}\sqrt{\gamma{}E/a}+|\sigma^\prime_0|$, where $E$ is a representative elastic modulus of the packing and $a$ is the vertical extent of the growing pathway; this result is adapted from classical linear-elastic fracture mechanics, with interfacial tension playing the role of surface energy. Lastly, the capillary entry pressure is $p_c^e\sim{}4\gamma/d$. 

We expect fluidization under weak confinement because $p_c^y$ will always be the smallest threshold pressure for sufficiently small $|\sigma^\prime_0|$ since $p_c^y\propto|\sigma^\prime_0|$. We expect a transition to pathway opening as confinement increases because, although both $p_c^y$ and $p_c^\mathrm{frac}$ increase with $|\sigma^\prime_0|$, the former will increase faster since $\alpha>0$. Note that $p_c^\mathrm{frac}$ also depends on bubble size, and will always exceed $p_c^y$ while the bubble is small; however, we expect a transition to pathway opening at a critical cavity/pathway size $a^\star=\gamma{}E/(\alpha|\sigma^\prime_0|)^2$ that rapidly drops below the particle size as $|\sigma^\prime_0|$ increases. Finally, we expect a transition to pore invasion under strong confinement because $p_c^e$ is the least sensitive to $|\sigma^\prime_0|$, increasing slowly as pore throats are squeezed. In all three cases, buoyancy accelerates gas invasion because the capillary pressure at the top of the connected gas region increases linearly with $a$. The gas will begin to disconnect and rise due buoyancy alone when it reaches a height where $\Delta{\rho}ga$ exceeds the minimum threshold pressure.

\begin{figure}
    \centering
    \includegraphics[width=17.2cm]{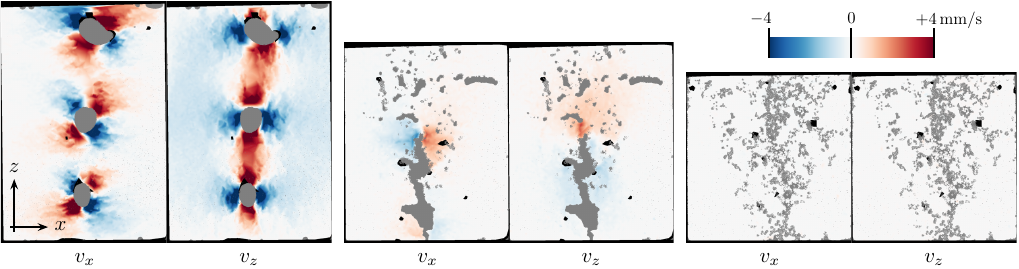}
    \caption{Horizontal ($v_x$) and vertical ($v_z$) components of the instantaneous particle velocity field for the snapshots shown in Fig.~\ref{fig:experiments}b, as calculated from particle tracking. Gray patches indicate gas. Fluidization (left two) involves yielding the packing, and is characterized by the flow of grains with and around the rising bubbles. Pathway opening (middle two) involves opening a narrow channel in the packing by pushing grains laterally apart, and is characterized by grain displacement primarily around the tip of the advancing pathway. In pore invasion (right two), the grains are essentially immobile.  \label{fig:mechanisms} }
\end{figure}

The qualitative distinction between these migration mechanisms is clear from the resulting air patterns (Fig.~\ref{fig:experiments}b), but also from the grain-scale mechanics that can be extracted from our high-resolution images. Figure~\ref{fig:mechanisms} shows the horizontal ($v_x$) and vertical ($v_z$) components of the instantaneous particle velocity field for the snapshots shown Fig.~\ref{fig:experiments}(b): $\phi^0_s = 0.49$ (fluidization), $\phi^0_s=0.59$ (pathway opening) and $\phi^0_s=0.69$ (pore invasion). In fluidization (leftmost two panels), particles rearrange freely as they rise with and then fall around the air bubbles, mimicking the flow field around a bubble rising in a liquid~\cite{maxworthy1986}. Pathways develop as confinement increases ($\phi^0_s = 0.59$, middle two panels), with grain motion increasingly reduced and localized around the advancing tip of pathway, much like the opening of a fracture in an elastic solid. In pore invasion (rightmost two panels), the particles are nearly motionless and the packing acts as a rigid porous medium.

\subsection{Emergence of mixed migration regimes}\label{ss:mixed}

The three distinct migration regimes described above are well known from previous work~\cite[\textit{e.g.},][]{jain2009,holtzman2010,sandnes2011}. However, the transitions between these regimes remain relatively unexplored, in part because no previous study has been able to capture all three in the same experimental system. We traverse the full range of behaviors here by controlling $\phi^0_s$, allowing us to explore these transitions in detail.

\begin{figure*}
    \centering
    \includegraphics[width=17.2cm]{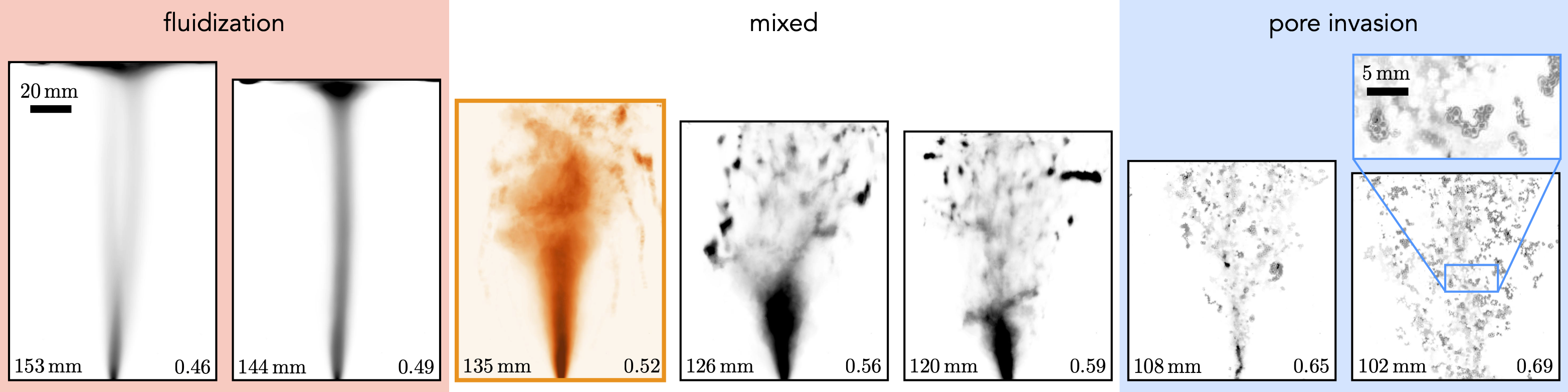}
    \caption{Air-occupancy images collapse an ensemble of episodic migration events into a single fingerprint. The darkness of each pixel is proportional to its average air saturation over the entire experiment, such that persistent trajectories and trapped blobs are darker, whereas ephemeral features are lighter. As $\phi_s^0$ increases from $0.46$ to $0.69$ (left to right), fluidization (vertical streaks) transitions to pathway opening (blurred fans with streaks and dots) and then to pore invasion (particle-scale spattering). The orange panel ($\phi_s^0=0.52$) corresponds to the high-capture mode, as discussed in \S~\ref{ss:high-capture}. \label{fig:occupancy} }
\end{figure*}

All three regimes exhibit episodic invasion, trapping, and venting events that occur on a timescale of seconds, repeating 10\,s or 100\,s of times during a given 10-minute experiment. Hence, for a qualitative, time-averaged view of migration mechanisms, we collapse each experiment into a single ``occupancy'' image in which the darkness of each pixel is proportional to its average air saturation over the entire experiment; accordingly, persistent or recurring features are darker, whereas ephemeral features are lighter~(Fig.~\ref{fig:occupancy}). These occupancy images serve as useful visual markers to track migration regimes in a time-averaged sense.

In fluidization, air migrates in discrete bubbles that rise vertically; the associated occupancy images show one or two smooth vertical streaks that end in a small darker region at the piston, where bubbles linger before venting (Fig.~\ref{fig:occupancy}, $\phi_s^0\lesssim{}0.50$). In pore invasion, pores drain and refill as pore-scale pathways connect, disconnect, and reconnect. The associated occupancy images show a low-intensity, grain-scale ``spattering'' with fine-scale particle outlines and a macroscopic V-shape (Fig.~\ref{fig:occupancy}, $\phi_s^0\gtrsim{}0.65$). The smooth transition between these two limits spans a broad range of intermediate solid fractions (\textit{i.e.}, $0.51\lesssim\phi_s^0\lesssim0.64$), and is characterized by the emergence of \textit{mixed} migration regimes. Specifically, air migration occurs via a different mechanism in the upper part of the flow cell than in the lower part. As $\phi_s^0$ increases, pathway opening first emerges at the top of the flow cell. For example, the occupancy image for $\phi_s^0=0.52$ illustrates the co-existence of fluidization and pathway opening, with a dark vertical streak near the bottom (signature of fluidization) that broadens and then transitions into an irregular fan of faint spots and streaks in the upper half of the flow cell (signature of pathway opening). This irregular fan lacks a single coherent streak because subsequent pathways do not consistently follow the same trajectory. The packing retains essentially no memory of previous pathways because pathway opening is associated with limited or no rearrangement of the particles, and pathways ``heal'' completely when they close because the particles are non-cohesive. Pathway opening is also characterized by the emergence of trapping, where blobs of air become disconnected from the main pathway and remain nearly stationary for long periods of time (isolated dark blobs in the occupancy images). As $\phi_s^0$ increases, the vertical transition between fluidization and pathway opening shifts downward until, at $\phi_s^{0,\mathrm{frac}}\approx0.58$, fluidization is completely suppressed and air appears to enter the flow cell via pathway opening. Overlapping with this transition, we see the emergence and expansion of pore invasion at the top of the flow cell from around $\phi_s^0\approx0.55$. Pathway opening vanishes as pore invasion reaches the bottom of the flow cell at $\phi_s^{0,\mathrm{rigid}}\approx 0.65$.

We previously discussed migration regimes in the context of $\phi_s^0$; clearly, however, the migration regime should depend locally on the actual value of $\phi_s$. The coexistence of different migration regimes in the flow cell is the signature of local variations in $\phi_s$. The fact that we see two distinct regimes, one in the bottom part of the flow cell and one in the top part, suggests that $\phi_s$ varies systematically in the vertical direction. To rationalize this observation, consider the response of the packing to the injection of air. The total volume of the flow cell is fixed, as is the total volume of particles within the flow cell. As a result, the addition of air can only be accommodated by a corresponding reduction in the volume of liquid, and liquid can only leave via the semi-permeable piston at the top. In pore invasion, the invasion of gas forces a net upward motion of liquid, which flows upward through the pore space and out via the piston. In fluidization and pathway opening, the invasion of gas forces a net upward motion of both liquid and solid. In the lower part of the flow cell, the liquid and the solid travel together because the gas can only separate the liquid from the solid by entering the pore space; as a result, the solid fraction remains roughly equal to $\phi_s^0$. In the upper part of the flow cell, the compression of the packing against the permeable piston leads to the formation of a compaction layer where the solid fraction increases as liquid is squeezed out. The result is a classical consolidation scenario~\citep[\textit{e.g.},][]{hewitt-physfluids-2016, macminn-prapplied-2016}.

The thickness of the compaction layer is set by the balance between the elasticity of the packing, which resists compression, and the viscous resistance to flow of liquid through the pore space. If viscosity dominates, the compaction layer will be localized near the piston; if elasticity dominates, the compaction layer will span the entire flow cell and $\phi_s$ will be uniform and greater than $\phi_s^0$. We transition from the former to the latter as $\phi_s^0$ increases. Thus, gas invasion in cavities or pathways increases the mean value of $\phi_s$ by generating a compaction layer in the upper part of the flow cell, which can itself then trigger an internal transition in migration regime.

\subsection{From mixed migration regimes to gas trapping}\label{ss:high-capture}

To connect air migration patterns more quantitatively to invasion, trapping, and venting, we consider the injection capillary pressure $p_c^\mathrm{inj}(t)$ and the total volume of air inside the flow cell $V_\mathrm{air}(t)$. We measure the latter via image processing (standard thresholding). We plot these quantities against time in Figures~\ref{fig:trapping}(a) and (c), respectively, for the three representative experiments shown in Figure~\ref{fig:mechanisms}. In all cases, $p_c^\mathrm{inj}$ and $V_\mathrm{air}$ reach a quasi-steady state in which they fluctuate in time with some characteristic amplitude and frequency around a mean value.

Fluctuations in $V_\mathrm{air}$ are due in part to the episodic nature of air invasion into the flow cell, which is itself related to the buildup and release of pressure during injection at constant nominal $Q$: $p_c^\mathrm{inj}$ increases until it exceeds the minimum threshold pressure (see \S\ref{ss:regimes}), after which invasion begins and continues for some period of time as $p_c^\mathrm{inj}$ drops~\cite{sandnes2011}. The amount of confinement has a strong impact on the amplitude and frequency of these pressure fluctuations, with the average peak value $\langle{}\mathrm{max}(p_c^\mathrm{inj})\rangle$ increasing relatively steadily and monotonically with $\phi_s^0$ (Fig.~\ref{fig:trapping}b).

Confinement has an even stronger impact on the amplitude and frequency of fluctuations in $V_\mathrm{air}$. Unlike with $p_c^\mathrm{inj}$, however, confinement also has a strong and non-monotonic impact on the mean air volume $\langle{}V_\mathrm{air}\rangle$ (Fig.~\ref{fig:trapping}d). For small $\phi_s^0$ (fluidization), $\langle{}V_\mathrm{air}\rangle$ increases weakly with $\phi_s^0$, hovering close to 0.5~mL. $\langle{}V_\mathrm{air}\rangle$ increases sharply at around $\phi_s^0 = 0.52$, which coincides with the emergence of pathway opening at the top of the flow cell (orange panel in Fig.~\ref{fig:occupancy}). Thereafter, $\langle{}V_\mathrm{air}\rangle$ decreases steadily with $\phi_s^0$ through the mixed regimes and into pure pore invasion. We decompose $\langle{}V_\mathrm{air}\rangle$ into the separate contributions from air in macroscopic cavities in the packing, $\langle{}V_\mathrm{open}\rangle$, and from air in the pore space of the packing, $\langle{}V_\mathrm{pore}\rangle$, where $\langle{}V_\mathrm{air}\rangle=\langle{}V_\mathrm{open}\rangle+\langle{}V_\mathrm{pore}\rangle$. The jump in $\langle{}V_\mathrm{air}\rangle$ at around $\phi_s^0 = 0.52$ is due entirely to a jump in $\langle{}V_\mathrm{open}\rangle$, which then decreases smoothly toward zero as the growing confinement suppresses the opening of bubbles and pathways. Meanwhile, $\langle{}V_\mathrm{pore}\rangle$ begins increasing gently as pore invasion emerges, and eventually dominates for $\phi_s^0\gtrsim{}0.65$.

\begin{figure*}
    \centering
    \includegraphics[width=15cm]{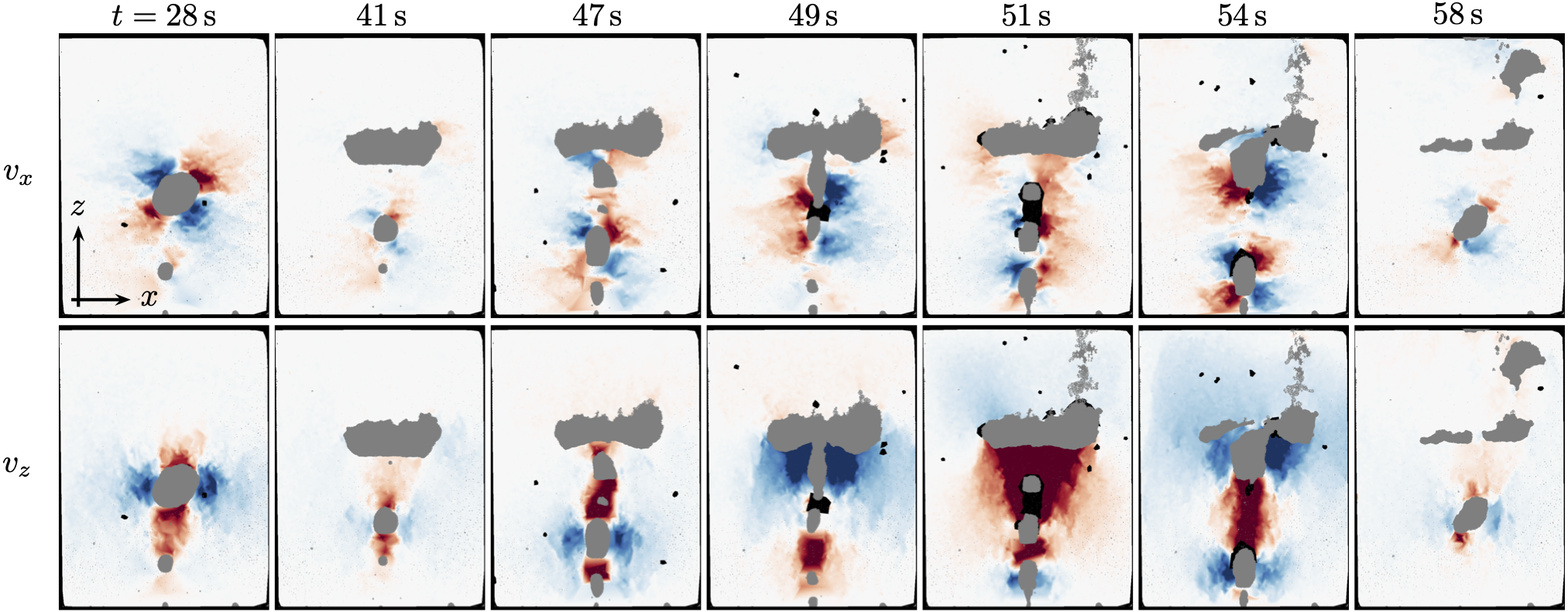}
    \caption{Time evolution of the horizontal ($v_x$, top row) and vertical ($v_z$, bottom row) components of the instantaneous particle velocity field for $\phi_s^0=0.52$ (orange panel in Fig.~\ref{fig:occupancy}), as calculated from particle tracking. The color scale is the same as in Figure~\ref{fig:mechanisms}. \label{fig:critical} }
\end{figure*}

The sharp increase in $\langle{}V_\mathrm{air}\rangle$ at around $\phi_s^0=0.52$ is a direct result of the fact that gas invasion leads to the formation of a compaction layer in the upper part of the flow cell. We illustrate the corresponding sequence of events in Figure~\ref{fig:critical} with images from the experiment at $\phi_s^0=0.52$ (orange panel in Fig.~\ref{fig:occupancy}), which is immediately after the jump in $\langle{}V_\mathrm{air}\rangle$. Air invades the flow cell as macroscopic bubbles that rise relatively easily via fluidization ($t=28\,\mathrm{s}$). However, the invasion of these bubbles at the bottom pushes the particle-liquid mixture upward, forming a compaction layer at the top. For this value of $\phi_s^0$, the thickness and/or solid fraction of the compaction layer is large enough to obstruct further bubble migration. The rigidification of the upper part of the packing is evident in the particle velocity fields, which show strong particle motion only in the lower part. Simultaneously, $\phi_s$ is still close to $\phi_s^0$ in the lower part of the flow cell and bubbles continue to enter and rise via fluidization. The result is that bubbles enter, rise, and then accumulate in a single, large suspended bubble ($t=41$--$51\,s$), and the volume of air in the flow cell steadily increases. This increase exacerbates the situation by increasing the thickness and/or solid fraction of the compaction layer, making it even more difficult for the suspended bubble to rise or vent.

The suspended bubble eventually vents when the capillary pressure at its top grows large enough to overcome the local threshold pressure (the minimum of $p_c^\mathrm{frac}(\phi_s)$ and $p_c^e$). Two mechanisms contribute to the growth in capillary pressure at the top of the suspended bubble: (1)~its height, which increases suddenly each time a rising bubble joins from below ($t=47$--$49\,\mathrm{s}$), and (2)~the effective stress in the surrounding packing, which increasingly squeezes the suspended bubble as more and more air enters the flow cell. Once the threshold capillary pressure is reached, the suspended bubble collapses as gas vents by either pathway opening or pore invasion ($t=49$--$58\,s$) and then the process repeats.

\begin{figure*}
    \centering
    \includegraphics[width=17.2cm]{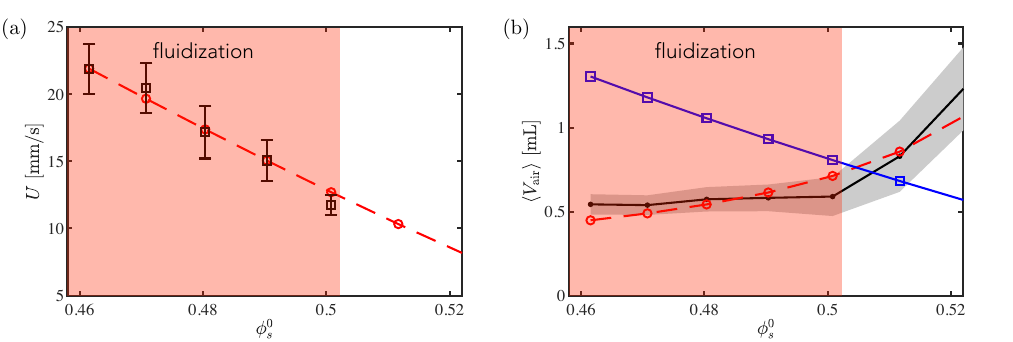}
    \caption{(a)~The characteristic bubble rise velocity $U$ as a function of $\phi_s^0$. The open squares correspond to the experimental data, while the dashed line is the result of our scaling law. (b) The plot of theoretical $\langle{}V_\mathrm{air}\rangle$ (open circles) is overlaid with experimental data, as a function of $\phi_s^0$.  For $\phi_s^0 < 0.5$, the model qualitatively replicates the gradual increase in $\langle{}V_\mathrm{air}\rangle$ with $\phi_s^0$. The plot also shows the minimum volume of air ($V_\mathrm{air}^\mathrm{req}$) required to rigidify the packing and to stop the rising bubble in blue squares. When the rigidified packing is assumed to have thickness $l_c = 4\,{\rm cm}$, $\langle{}V_\mathrm{air}\rangle{}^\mathrm{req}$ and theoretical $\langle{}V_\mathrm{air}\rangle$ match at $\phi_s^0 = 0.506$, which signifies the transition to enhanced gas trapping. \label{fig:fluidization} }
\end{figure*}

The process described above leads to a sharp increase in $\langle{}V_\mathrm{air}\rangle$ at the smallest value of $\phi_s^0$ that is large enough to trigger the transition to pathway opening at the top of the flow cell. For a slightly smaller value of $\phi_s^0$, bubbles would rise and escape without triggering this transition. We now consider the mechanics that leads to this critical value of $\phi_s^0$ in the context of a toy model. Note that the purpose of this model is not to predict the value of $\phi_s^0$ at which the jump will occur, but rather to rationalize the occurence of the jump.

In fluidization, $\langle{}V_\mathrm{air}\rangle$ can be estimated by considering the residence time of each bubble within the flow cell, and the residence time can be estimated from the rise velocity. We assume that the rise velocity $U$ of each bubble satisfies the scaling relation due to  \citet{maxworthy1986},
\begin{equation}\label{eq:U}
    U(\phi_s^0)\propto \frac{\rho_lb^2 g}{\mu_l(\phi_s^0)}.
\end{equation}
Note that $U$ is independent of the bubble radius $R_b$, and that this scaling is only appropriate when $R_b\gg b$ \cite{maxworthy1986}. The effective density of the mixture $\rho_l$ is independent of $\phi_s^0$ because the particles and the liquid have nearly the same density. The effective viscosity of the mixture $\mu_l(\phi_s^0)$ is a monotonically increasing function of $\phi_s^0$ that we estimate from the empirical correlation of \citet{morris1999} for viscous suspensions,
\begin{equation}
	\frac{\mu_l}{\mu_0} = 1+{5\over 2}\phi_s\left(1-\frac{\phi_s}{\phi_m}\right)^{-1}+0.1\left(\frac{\phi_s}{\phi_m}\right)^2\left(1-\frac{\phi_s}{\phi_m}\right)^{-2},
\end{equation}
where $\mu_0$ is the viscosity of the pure liquid and $\phi_m$ is the maximum allowable packing fraction. The latter is not well defined in our system, so we take it to be the value at which bubbles no longer enter via fluidization, $\phi_m= \phi_s^{0,\mathrm{frac}}\approx{}0.58$. Taking $\phi_s=\phi_s^0$ and setting the constant of proportionality in Eq.~\eqref{eq:U} to be $0.14$ provides a reasonable match with rise velocities extracted from our fluidization experiments (Fig.~\ref{fig:fluidization}a).

We estimate the average volume of air inside the cell as $\langle{}V_\mathrm{air}\rangle\approx [hf/U(\phi_s^0)]\pi R_b^2 b$, where $f$ is the frequency of bubble entry and the quantity $hf/U(\phi_s^0)$ is then the average number of bubbles inside the cell at any time. Taking $f\approx 0.2\,{\rm s}^{-1}$ and $R_b\approx 1\,{\rm cm}$, as suggested by our experimental observations, we find that this simple model does a reasonable job of capturing the gradual increase in $\langle{}V_\mathrm{air}\rangle$ with $\phi_s^0$ for $\phi_s^0 < 0.5$ (Fig.~\ref{fig:fluidization}b, red circles). Note that we have ignored the impact of the compaction layer on bubble velocity, as well as any impact of $\phi_s^0$ on $f$ and $R_b$. The observed decrease of $f$ with $\phi_s^0$ (Fig.~\ref{fig:trapping}c) would contribute to our predicted $\langle{}V_\mathrm{air}\rangle(\phi_s^0)$ being somewhat steeper than the experimental measurements.

In our experiments, a further increase in $\phi_s^0$ above $\phi_s^0 \approx 0.5$ leads to a sharp increase in $\langle{}V_\mathrm{air}\rangle$. As discussed above, this jump results from the fact that air invasion generates a compaction layer at the top of the flow cell. To capture this idea, we suppose that the packing consists of two discrete regions: A dense compaction layer where $\phi_s\approx{}\phi_s^\mathrm{frac}$ and a fluidized lower region where $\phi_s\approx{}\phi_s^0$. The time-averaged thickness $\langle{}l\rangle$ of the compaction layer is readily calculated in terms of $\phi_s^0$, $\phi_s^\mathrm{frac}$, $\langle{}V_\mathrm{air}\rangle{}$, and the cell geometry via mass conservation, $\langle{}l\rangle{}=\phi_s^0\langle{}V_\mathrm{air}\rangle{}/[(\phi_s^\mathrm{frac} - \phi_s^0)wb]$. Supposing that a critical layer thickness $\langle{}l\rangle{}_c$ is required to obstruct a rising bubble, the mean air volume that would be required for $\langle{}l\rangle{}$ to reach $\langle{}l\rangle{}_c$ is then
\begin{align}\label{eq:Vair_req}
    \langle{}V_\mathrm{air}\rangle{}^\mathrm{req} = \frac{(\phi_s^\mathrm{frac} - \phi_s^0)\langle{}l\rangle{}_cwb}{\phi_s^0},
\end{align}
which is a decreasing function of $\phi_s^0$. Taking $\langle{}l\rangle{}_c$ to be roughly the distance between the top of the suspended bubble and the top of the flow cell, Figure~\ref{fig:critical} suggests that $\langle{}l\rangle{}_c\approx{}4\,{\rm cm}$. Figure~\ref{fig:fluidization}(b) shows that, for $\langle{}l\rangle{}_c=4\,{\rm cm}$, our estimate of $\langle{}V_\mathrm{air}\rangle{}^\mathrm{req}$ (blue squares) intersects our estimate of $\langle{}V_\mathrm{air}\rangle{}$ (red circles) at $\phi_s^0 = 0.506$, which would then be the trigger for a jump in $\langle{}V_\mathrm{air}\rangle$ at around the same value of $\phi_s^0$ as in our experiments. Note that the fluid and solid mechanics that control the precise value of $\langle{}l\rangle{}_c$ remain unclear; varying $\langle{}l\rangle{}_c$ from $1$ to $5\,{\rm cm}$ produces a transitional value of $\phi_s^0$ that ranges from $0.439$ to $0.513$. Clearly, however, the volume of air needed to generate a compaction layer that obstructs bubble migration (\textit{i.e.},~$\langle{}V_\mathrm{air}\rangle{}^\mathrm{req}$) should decrease with $\phi_s^0$, whereas the actual mean volume of air in the flow cell (\textit{i.e.},~$\langle{}V_\mathrm{air}\rangle$) should increase with $\phi_s^0$. We expect a jump in $\langle{}V_\mathrm{air}\rangle$ around the value of $\phi_s^0$ at which the two volumes cross.

\section{Summary and Discussion}

We have experimentally investigated the mechanisms of air migration, trapping, and venting in a soft, liquid-saturated granular material in a vertical Hele-Shaw cell. Our granular material is a quasi-2D packing of hydrogel particles, which allows for high-resolution visualization of air distributions and particle motion, providing an ideal platform for connecting particle-scale flow and mechanics with macroscopic air invasion, trapping, and venting. This system also allows us to span the full range of expected migration mechanisms by varying the initial solid fraction over a wide range. Similar experiments with rigid, frictional particles would most likely exhibit a much narrower transition from fluidization to pore invasion since they allow for limited and very short-range rearrangements once jammed.

We have shown that, as the confining pre-stress increases, air migration transitions smoothly from fluidization to pore invasion via a series of intermediate mixed regimes that combine fluidization with pathway opening or pathway opening with pore invasion. In our system, the emergence of mixed migration regimes is rooted in the spontaneous formation of a compaction layer at the top of the flow cell. The compaction layer forms because macroscopic air bubbles and pathways reduce the amount of space available to the particles, and it forms at the top of the flow cell because that is the only place that fluid can leave. The formation of the compaction layer results in a situation in which it is substantially easier for air to enter the flow cell at the bottom than it is for air to exit the flow cell at the top, which then leads to a sharp increase in the volume of air in the flow cell. These mixed regimes thus enable an anomalously large amount of gas trapping, as well as anomalously large venting events. In the end-member regimes of fluidization and pore invasion, in contrast, the volume of air in the flow cell is much smaller and depends comparatively weakly on $\phi_s^0$, because the packing is either sufficiently ``dilute'' (in fluidization) or sufficiently rigid (in pore invasion) that the resistance to air migration is nearly uniform within the flow cell.

Mixed migration regimes have also been observed in unconfined vertical packings of dense particles, in which the confining stress is zero at the free surface and increases roughly linearly with depth due to the negative buoyancy of the particles. In those cases, the mixed regimes are inverted relative to our experiments, with a transition from pore invasion to pathway opening or from pathway opening to fluidisation as the gas rises~\citep{varas2011a, varas2011b, varas2013, varas2015}. Air invasion, trapping, and venting have not yet been quantified in that context.

Our results suggest that the relationship between invasion, trapping, and venting could be controlled by modulating the confining stress, which could be useful in the design of bio-reactors or fluidized-bed chemical reactors. We plan to explore this idea in future work. The results presented here will also play a valuable role in informing the development of a continuum model for this three-phase poromechanical flow problem; such a model would itself have broad applications.

\begin{acknowledgments}
This research was supported by the Royal Society under the International Exchanges Scheme (IE150885), by the European Research Council (ERC) under the European Union's Horizon 2020 Programme (Grant agreement No. 805469), and by the European Union under the Erasmus+ Program (to RL). The authors also acknowledge support from the John Fell OUP Research Fund (132/012) and the Maurice Lubbock Memorial Fund. The authors thank Rob Style for helpful discussions about fracture mechanics.
\end{acknowledgments}




%


\end{document}